\documentclass[aps,10pt,prl,twocolumn,floatfix,nobibnotes,superscriptaddress]{revtex4-2}
\usepackage{dcolumn,amsmath}
\usepackage{booktabs}
\usepackage{bm}
\usepackage{hyperref}
\usepackage{xcolor}             
\usepackage[utf8]{inputenc}
\usepackage{graphicx}
\usepackage{dcolumn}
\usepackage{bm}
\usepackage{amsmath}
\usepackage{color} 
\usepackage{soul}
\usepackage[normalem]{ulem}
\usepackage{hyperref}
\usepackage[english]{babel}
\usepackage{physics}
\usepackage{tabularx}
\usepackage{upgreek}

\newcounter{firstbib}
\setcitestyle{super}
\usepackage{etoolbox}
\patchcmd{\section}
  {\centering}
  {\raggedright}
  {}
  {}
\patchcmd{\subsection}
  {\centering}
  {\raggedright}
  {}
  {}
\usepackage[normalem]{ulem}

\makeatletter
\renewcommand{\fnum@figure}{Fig.~\thefigure}
\renewcommand{\fnum@table}{Table~\thetable}
\def\@hangfrom@section#1#2#3{\@hangfrom{#1#2}#3}
\def\@hangfroms@section#1#2{#1#2}
\renewcommand\section{\@startsection {section}{1}{0pt}
  {-3.5ex \@plus -1ex \@minus -.2ex}
  {2.3ex \@plus.2ex}
  {\normalfont\large\bfseries}}
\makeatother

\makeatletter
\let\doauthor@orig\doauthor

\def\doauthor#1#2#3{%
  \ignorespaces#1\unskip\@listcomma
  \begingroup
    #3%
  \@if@empty{#2}{%
    \endgroup{}{}%
  }{%
    \endgroup{\comma@space}{}%
    \frontmatter@footnote{#2}%
    \def\tempauthor{#1}%
    \def\targetauthor{Wenhao Bu}%
    \ifx\tempauthor\targetauthor
  \textsuperscript{,\,}%
  \frontmatter@footnote{Corresponding author: eardius@163.com}%
\fi
  }%
  \space \@listand
}%
\makeatother

\begin{document}
\newcommand{\equalcontrib}{\ensuremath{^{*}}}
\newcommand{\ThTitle}{$^{229}$Th}
\pdfstringdefDisableCommands{\def\ThTitle{229Th}}

\affiliation{State Key Laboratory of Low-Dimensional Quantum Physics, Department of Physics, Tsinghua University, Beijing 100084, China}

\affiliation{Beijing Academy of Quantum Information Sciences, Beijing 100193, China}

\affiliation{Research Center of Laser Crystal, Advanced Laser and Optoelectronic Functional Materials Department,\linebreak
\scalebox{1}[1]{Shanghai Institute of Optics and Fine Mechanics, Chinese Academy of Sciences, Shanghai 201800, China}}

\affiliation{State Key Laboratory of Functional Crystals and Devices, Shanghai Institute of Ceramics, Chinese Academy of Sciences, Shanghai 201899, China}

\affiliation{Center of Materials Science and Optoelectronics Engineering, University of Chinese Academy of Sciences, Beijing 100049, China}

\author{Beichen Huang}
\thanks{These authors contributed equally to this work.}
\affiliation{State Key Laboratory of Low-Dimensional Quantum Physics, Department of Physics, Tsinghua University, Beijing 100084, China}
\affiliation{Beijing Academy of Quantum Information Sciences, Beijing 100193, China}

\author{Gaowei Yan}
\thanks{These authors contributed equally to this work.}
\affiliation{State Key Laboratory of Low-Dimensional Quantum Physics, Department of Physics, Tsinghua University, Beijing 100084, China}
\affiliation{Beijing Academy of Quantum Information Sciences, Beijing 100193, China}

\author{Qi Xiao}
\thanks{These authors contributed equally to this work.}
\affiliation{State Key Laboratory of Low-Dimensional Quantum Physics, Department of Physics, Tsinghua University, Beijing 100084, China}
\affiliation{Beijing Academy of Quantum Information Sciences, Beijing 100193, China}

\author{Wenhao Bu}
\thanks{These authors contributed equally to this work.}
\affiliation{Beijing Academy of Quantum Information Sciences, Beijing 100193, China}
\affiliation{State Key Laboratory of Low-Dimensional Quantum Physics, Department of Physics, Tsinghua University, Beijing 100084, China}

\author{Chengchun Zhao}
\thanks{These authors contributed equally to this work.}
\affiliation{Research Center of Laser Crystal, Advanced Laser and Optoelectronic Functional Materials Department,\linebreak
\scalebox{1}[1]{Shanghai Institute of Optics and Fine Mechanics, Chinese Academy of Sciences, Shanghai 201800, China}}
\affiliation{Center of Materials Science and Optoelectronics Engineering, University of Chinese Academy of Sciences, Beijing 100049, China}

\author{Zhen Zhang}
\thanks{These authors contributed equally to this work.}
\affiliation{State Key Laboratory of Functional Crystals and Devices, Shanghai Institute of Ceramics, Chinese Academy of Sciences, Shanghai 201899, China}

\author{Chao Yan}
\thanks{These authors contributed equally to this work.}
\affiliation{State Key Laboratory of Thorium Energy, Shanghai Institute of Applied Physics, Chinese Academy of Sciences, Shanghai 201800, China}

\author{Zhi-Ang Chen}
\affiliation{International Center for Quantum Materials, School of Physics, Peking University, Beijing 100871, China}
\affiliation{Beijing Key Laboratory of Quantum Devices, Peking University, Beijing 100871, China}

\author{Peixiong Zhang}
\affiliation{Research Center of Laser Crystal, Advanced Laser and Optoelectronic Functional Materials Department,\linebreak
\scalebox{1}[1]{Shanghai Institute of Optics and Fine Mechanics, Chinese Academy of Sciences, Shanghai 201800, China}}

\author{Gleb Penyazkov}
\affiliation{State Key Laboratory of Low-Dimensional Quantum Physics, Department of Physics, Tsinghua University, Beijing 100084, China}

\author{Zhenhai Zhan}
\affiliation{State Key Laboratory of Low-Dimensional Quantum Physics, Department of Physics, Tsinghua University, Beijing 100084, China}

\author{Lingfeng Yan}
\affiliation{State Key Laboratory of Low-Dimensional Quantum Physics, Department of Physics, Tsinghua University, Beijing 100084, China}

\author{Yuefei Wang}
\affiliation{Beijing Academy of Quantum Information Sciences, Beijing 100193, China}

\author{Lin Li}
\affiliation{Research Center of Laser Crystal, Advanced Laser and Optoelectronic Functional Materials Department,\linebreak
\scalebox{1}[1]{Shanghai Institute of Optics and Fine Mechanics, Chinese Academy of Sciences, Shanghai 201800, China}}

\author{Shanming Li}
\affiliation{Research Center of Laser Crystal, Advanced Laser and Optoelectronic Functional Materials Department,\linebreak
\scalebox{1}[1]{Shanghai Institute of Optics and Fine Mechanics, Chinese Academy of Sciences, Shanghai 201800, China}}

\author{Dapeng Jiang}
\affiliation{State Key Laboratory of Functional Crystals and Devices, Shanghai Institute of Ceramics, Chinese Academy of Sciences, Shanghai 201899, China}

\author{Xiaobo Qian}
\affiliation{State Key Laboratory of Functional Crystals and Devices, Shanghai Institute of Ceramics, Chinese Academy of Sciences, Shanghai 201899, China}

\author{Xuegang Liu}
\affiliation{Institute of Nuclear and New Energy Technology, Tsinghua University, Beijing 100084, China}

\author{Qiange He}
\affiliation{Institute of Nuclear and New Energy Technology, Tsinghua University, Beijing 100084, China}

\author{Taoxiang Sun}
\affiliation{Institute of Nuclear and New Energy Technology, Tsinghua University, Beijing 100084, China}

\author{Haochen Tian}
\affiliation{Division of Time and Frequency Metrology, National Institute of Metrology, Beijing 100029, China}
\affiliation{Hefei National Laboratory, Hefei 230088, China}

\author{Bingkun Lu}
\affiliation{Division of Time and Frequency Metrology, National Institute of Metrology, Beijing 100029, China}

\author{Ningyuan Ma}
\affiliation{State Key Laboratory of Low-Dimensional Quantum Physics, Department of Physics, Tsinghua University, Beijing 100084, China}

\author{Juxian Li}
\affiliation{State Key Laboratory of Low-Dimensional Quantum Physics, Department of Physics, Tsinghua University, Beijing 100084, China}

\author{Yanzhang Wu}
\affiliation{State Key Laboratory of Low-Dimensional Quantum Physics, Department of Physics, Tsinghua University, Beijing 100084, China}

\author{Qiaorui Gong}
\affiliation{Research Center of Laser Crystal, Advanced Laser and Optoelectronic Functional Materials Department,\linebreak
\scalebox{1}[1]{Shanghai Institute of Optics and Fine Mechanics, Chinese Academy of Sciences, Shanghai 201800, China}}
\affiliation{Center of Materials Science and Optoelectronics Engineering, University of Chinese Academy of Sciences, Beijing 100049, China}

\author{Yuxiang Li}
\affiliation{Research Center of Laser Crystal, Advanced Laser and Optoelectronic Functional Materials Department,\linebreak
\scalebox{1}[1]{Shanghai Institute of Optics and Fine Mechanics, Chinese Academy of Sciences, Shanghai 201800, China}}
\affiliation{Center of Materials Science and Optoelectronics Engineering, University of Chinese Academy of Sciences, Beijing 100049, China}

\author{Haoyu Shi}
\affiliation{State Key Laboratory of Low-Dimensional Quantum Physics, Department of Physics, Tsinghua University, Beijing 100084, China}

\author{Xiangliang Li}
\affiliation{Beijing Academy of Quantum Information Sciences, Beijing 100193, China}

\author{Longsheng Ma}
\affiliation{State Key Laboratory of Precision
Spectroscopy, East China Normal University, Shanghai
200062, China}

\author{Shining Zhu}
\affiliation{National Laboratory of Solid-State
Microstructures, School of Physics, Collaborative Innovation
Center of Advanced Microstructures, Nanjing University,
Nanjing, Jiangsu 210093, China}

\author{Yuxiang Mo}
\affiliation{State Key Laboratory of Low-Dimensional Quantum Physics, Department of Physics, Tsinghua University, Beijing 100084, China}

\author{Jun Lin}
\affiliation{State Key Laboratory of Thorium Energy, Shanghai Institute of Applied Physics, Chinese Academy of Sciences, Shanghai 201800, China}

\author{Li You}
\affiliation{State Key Laboratory of Low-Dimensional Quantum Physics, Department of Physics, Tsinghua University, Beijing 100084, China}
\affiliation{Beijing Academy of Quantum Information Sciences, Beijing 100193, China}
\affiliation{Frontier Science Center for Quantum Information, Beijing 100084, China}
\affiliation{Hefei National Laboratory, Hefei 230088, China}

\author{Yige Lin}
\affiliation{Division of Time and Frequency Metrology, National Institute of Metrology, Beijing 100029, China}
\affiliation{Hefei National Laboratory, Hefei 230088, China}

\author{Xibo Zhang}
\affiliation{International Center for Quantum Materials, School of Physics, Peking University, Beijing 100871, China}
\affiliation{Beijing Key Laboratory of Quantum Devices, Peking University, Beijing 100871, China}
\affiliation{Hefei National Laboratory, Hefei 230088, China}

\author{Yin Hang}
\thanks{Corresponding author: yhang@siom.ac.cn}
\affiliation{Research Center of Laser Crystal, Advanced Laser and Optoelectronic Functional Materials Department,\linebreak
\scalebox{1}[1]{Shanghai Institute of Optics and Fine Mechanics, Chinese Academy of Sciences, Shanghai 201800, China}}
\affiliation{Center of Materials Science and Optoelectronics Engineering, University of Chinese Academy of Sciences, Beijing 100049, China}

\author{Liangbi Su}
\thanks{Corresponding author: suliangbi@mail.sic.ac.cn}
\affiliation{State Key Laboratory of Functional Crystals and Devices, Shanghai Institute of Ceramics, Chinese Academy of Sciences, Shanghai 201899, China}

\author{Shiqian Ding}
\thanks{Corresponding author: dingshq@mail.tsinghua.edu.cn}
\affiliation{State Key Laboratory of Low-Dimensional Quantum Physics, Department of Physics, Tsinghua University, Beijing 100084, China}
\affiliation{Beijing Academy of Quantum Information Sciences, Beijing 100193, China}
\affiliation{Frontier Science Center for Quantum Information, Beijing 100084, China}

\title{A nuclear clock synchronized to \(^{229}\)Th}

\begin{abstract}
\textbf{Atomic clocks have made time and frequency the most precisely measured quantities in physics, progressing from microwave standards that realize the SI second~\cite{MicrowaveClock} to optical clocks with unprecedented precision~\cite{2015RMP_Optical_atomic_clocks}.
A nuclear clock transfers the frequency reference from an electronic to a nuclear transition, and the uniquely low-lying, laser-accessible isomeric transition in $^{229}$Th currently offers the most practical route to compact, robust timekeeping and sensitive tests of fundamental physics~\cite{ClockPeik,CrystalHudson,ClockKuzmich,FlambaumVariation,Beeks2025NatCommunAlpha,ThoriumReviewNuClock}.
Realizing such a clock requires turning spectroscopy of the \(^{229}\)Th nuclear resonance~\cite{vonDerWense2016,2019EnergyMunich,2019EnergyJapan,2020Energy,2022EnergyCERN,LaserSpecSchumm,LaserSpecHudson,LaserSpecYe,FrequencyReproducibility2025} into a stable discriminator for steering a traceable oscillator.
Here we demonstrate a $^{229}$Th nuclear clock by stabilizing a continuous-wave, narrow-linewidth 148.4-nm vacuum-ultraviolet (VUV) laser~\cite{DingVUV2025} to a resolved, weakly temperature-sensitive nuclear transition~\cite{FrequencyReproducibility2025,Temperature} in $^{229}$Th$:$CaF$_2$ crystals~\cite{Beeks2023ThoriumCaF2,HangThCaF2,SuTGT}.
A $10$-$\mu\mathrm{W}$ VUV source generated by four-wave mixing in cadmium vapour~\cite{DingVUVproposal2024,Mat_elem,DingVUV2025} and phototube-based frequency-modulation absorption readout provide a fast, high-signal-to-noise nuclear discriminator.
The clock reaches a fractional frequency instability of $5\times10^{-13}/\sqrt{\tau/\mathrm{s}}$ for averaging time $\tau$. 
Clock-transition frequencies measured in two independently fabricated crystals agree at the \(10^{-13}\) level and are consistent with previous VUV-comb measurements on other \(^{229}\)Th:CaF\(_2\) crystals~\cite{FrequencyReproducibility2025}.
These results establish laser-addressed nuclei as operational clock references and provide a reproducible solid-state platform for compact nuclear clocks, nuclear quantum sensors and precision tests of fundamental physics.}
\end{abstract}

\date{\today}
\flushbottom
\maketitle

Laser generation and laser spectroscopy are rooted in the electronic structure of atoms, molecules and solids. Nuclear transitions, by contrast, usually occur at energies far above typical electronic transitions, leaving most nuclear resonances beyond direct coherent access with conventional optical spectroscopy. The $^{229}$Th isomer is a singular exception. Its transition energy lies in the vacuum-ultraviolet (VUV) range, making it the only known nuclear transition that can be directly addressed with current laser technology. This coincidence opens a route to quantum coherent control of a nucleus and, in particular, to a nuclear clock, in which the frequency reference is shifted from the electronic shell to the atomic nucleus.

Decades of work established the existence and energy of this low-lying state~\cite{1976Energy,1990Energy,1994Energy,2007Energy,vonDerWense2016,2019EnergyMunich,2019EnergyJapan,2020Energy,Thielking2018Nature,Th3JapanNature,ThoriumReviewPeik}. The observation of radiative decay from $^{229}$Th-doped crystals further showed that the isomer can remain long-lived in VUV-transparent hosts~\cite{2022EnergyCERN,QuenchingJapan,QuenchingPeik,QuenchingHudson,Guan2025XrayQuenching}.
Building on these developments, direct laser excitation of the $^{229}$Th isomer has recently been achieved~\cite{LaserSpecSchumm,LaserSpecHudson,LaserSpecYe,Hiraki2025LaserMossbauer}, and VUV-frequency-comb spectroscopy has enabled precision measurement of the nuclear transition frequency against optical atomic clocks~\cite{LaserSpecYe}. 
In crystalline hosts, spectroscopy has further resolved the solid-state structure of the nuclear resonance, including quadrupole-split transitions~\cite{LaserSpecYe}, site-dependent shifts~\cite{Hiraki2025LaserMossbauer,Perera2025HostOffsets}, temperature sensitivities~\cite{Temperature} and reproducible transition frequencies across crystals~\cite{FrequencyReproducibility2025}.
Together with the macroscopic number of nuclei provided by a solid-state host~\cite{ThF4,MossbauerHudson,Morgan2025APLNonlinearCrystals}, these advances open the possibility of compact and field-deployable nuclear frequency references.

Another enabling element for clock operation is a continuous-wave VUV source at the nuclear resonance. 
Resonantly enhanced four-wave mixing in metal vapour~\cite{DingVUV2025,DingVUVproposal2024,Mat_elem} and randomly quasi-phase-matched SrB\(_4\)O\(_7\)~\cite{SBO2016,PeikCWVUV} have provided routes to continuous-wave 148.4-nm radiation.
Converting nuclear laser spectroscopy into clock operation requires combining, in a single system, a traceable narrow-linewidth VUV probe, a $^{229}$Th-containing reference, and a high-bandwidth, high-signal-to-noise readout of weak nuclear response, together with a demonstration of nuclear-transition-frequency reproducibility in independently prepared references.

In this work, we develop and integrate all the ingredients required for $^{229}$Th clock operation.
We upgrade the cadmium-vapour four-wave-mixing source~\cite{DingVUV2025} to generate 10 $\mu$W of continuous-wave 148.4-nm radiation with an estimated sub-Hertz linewidth. 
We also fabricate, in a single growth run, a compact, high-quality $^{229}$Th:CaF$_2$ crystal from a solution containing only 1.4 $\mu$g (10 kBq) of $^{229}$Th. 
In fluorescence and absorption spectroscopy, we resolve a broad quadrupole-unresolved structure (QU) and a quadrupole-resolved structure (QR) comprising five narrow transitions, including line b, which is used for clock operation.
Combining this traceable high-power VUV probe with home-grown crystals, we demonstrate the operation of the first $^{229}$Th nuclear clock by using phototube-based frequency-modulation absorption spectroscopy to generate the nuclear discriminator and close a feedback loop to line b.
We then use the feedback-stabilized system to characterize the clock instability and test the reproducibility of the clock-transition frequency across two independently fabricated crystals and by comparison with previous VUV-comb measurements on other \(^{229}\)Th:CaF\(_2\) crystals~\cite{FrequencyReproducibility2025,ooi_2025_15751158}.

A near-simultaneous independent report has also demonstrated $^{229}$Th nuclear clock operation using the randomly quasi-phase-matched SrB$_4$O$_7$ VUV source, a highly doped $^{229}$Th:CaF$_2$ crystal and photomultiplier-tube-based absorption readout~\cite{PeikSchummClock}, building on related continuous-wave nuclear absorption spectroscopy with sub-nanowatt 148.4-nm light~\cite{Morawetz2026ContinuousCW}.

\begin{figure*}[htbp]
        \centering
    \includegraphics[width=\linewidth]{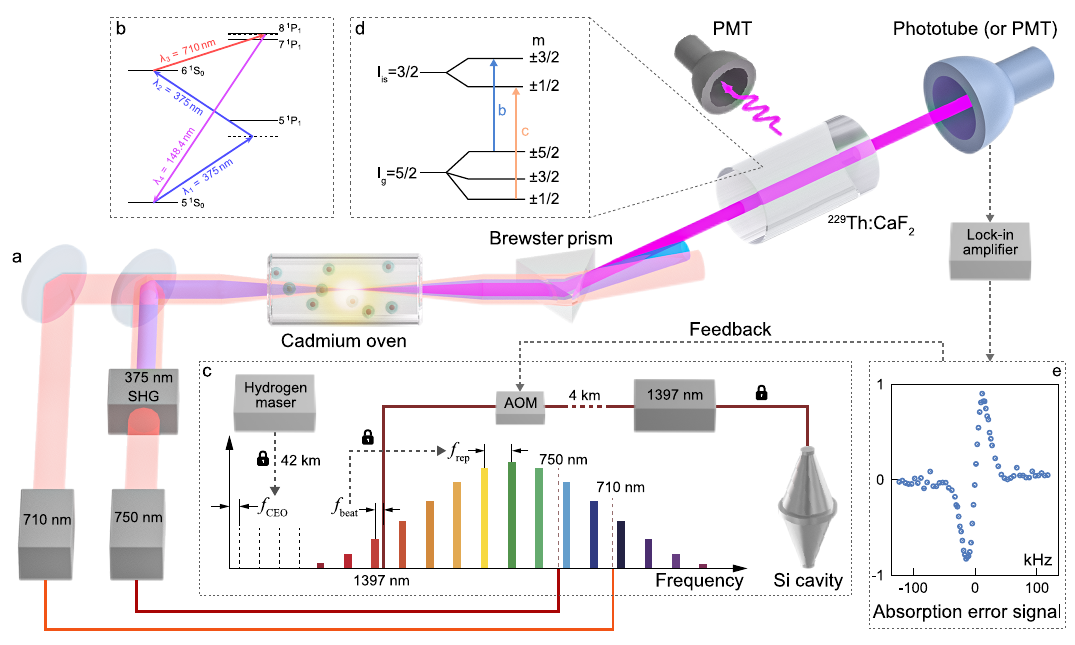}
    \caption{\textbf{Experimental platform for a solid-state $^{229}$Th nuclear clock.}
    \textbf{a,} Experimental setup. 
    Continuous-wave 148.4-nm VUV light is generated by four-wave mixing in cadmium vapour and sent to a home-grown $^{229}$Th:CaF$_2$ crystal. 
    Nuclear fluorescence is collected with a photomultiplier tube (PMT), while the transmitted VUV power is recorded with a phototube or a PMT. 
    During absorption spectroscopy and clock operation, the 710-nm laser is modulated, and the transmitted-VUV photocurrent is detected via lock-in demodulation. SHG, second-harmonic generation.
    \textbf{b,} Cadmium energy-level scheme for resonantly enhanced four-wave mixing. Two 375-nm photons drive the two-photon resonance between the $5 \,^1$S$_0$ and $6 \,^1$S$_0$ states, whereas one 710-nm photon completes the sum-frequency process to generate the 148.4-nm VUV field. 
    \textbf{c,} Optical frequency chain. 
    The 750-nm and 710-nm lasers are phase-locked to a self-referenced Er-fibre frequency comb, whose repetition rate ($f_{\mathrm{rep}}$) is stabilized by a 4-km-long phase-stabilized fibre link to a cryogenic-silicon-cavity-stabilized laser at 1,397~nm. The carrier-envelope-offset frequency ($f_{\mathrm{CEO}}$) is referenced to a hydrogen maser.
    \textbf{d,} {Quadrupole-split nuclear energy levels of $^{229}$Th in a CaF$_2$ host. Line b, a QR component corresponding to $m_{\mathrm{g}}=\pm5/2 \rightarrow m_{\mathrm{is}}=\pm3/2$, is used for clock operation. Line c is a QR component corresponding to \(m_g=\pm1/2\rightarrow m_{is}=\pm1/2\).}
    \textbf{e,} Absorption-based clock discriminator and feedback. The lock-in-demodulated absorption signal provides a dispersive error signal near line b, and the resulting frequency correction is applied to an AOM between the 1,397-nm reference laser and the frequency comb, thereby steering the VUV frequency onto the nuclear resonance. The plotted trace shows the normalized demodulated absorption signal measured with the optimized TS1 + phototube configuration. The frequency offset is referenced to the fitted line centre, and each data point corresponds to a 10-s acquisition. The zero crossing defines the clock lock point, and the local slope converts fluctuations of the normalized demodulated signal into residual detuning for the clock servo. The frequency-modulation amplitude for this trace is 12~kHz. Unless otherwise stated, all error bars throughout this work represent $1\sigma$ statistical uncertainties.}
    \label{fig:1}
\end{figure*}

\vspace{0.6cm}
\noindent \textbf{VUV laser for nuclear clock interrogation}

\begin{figure*}[htbp]
    \centering
    \includegraphics[width=\linewidth]{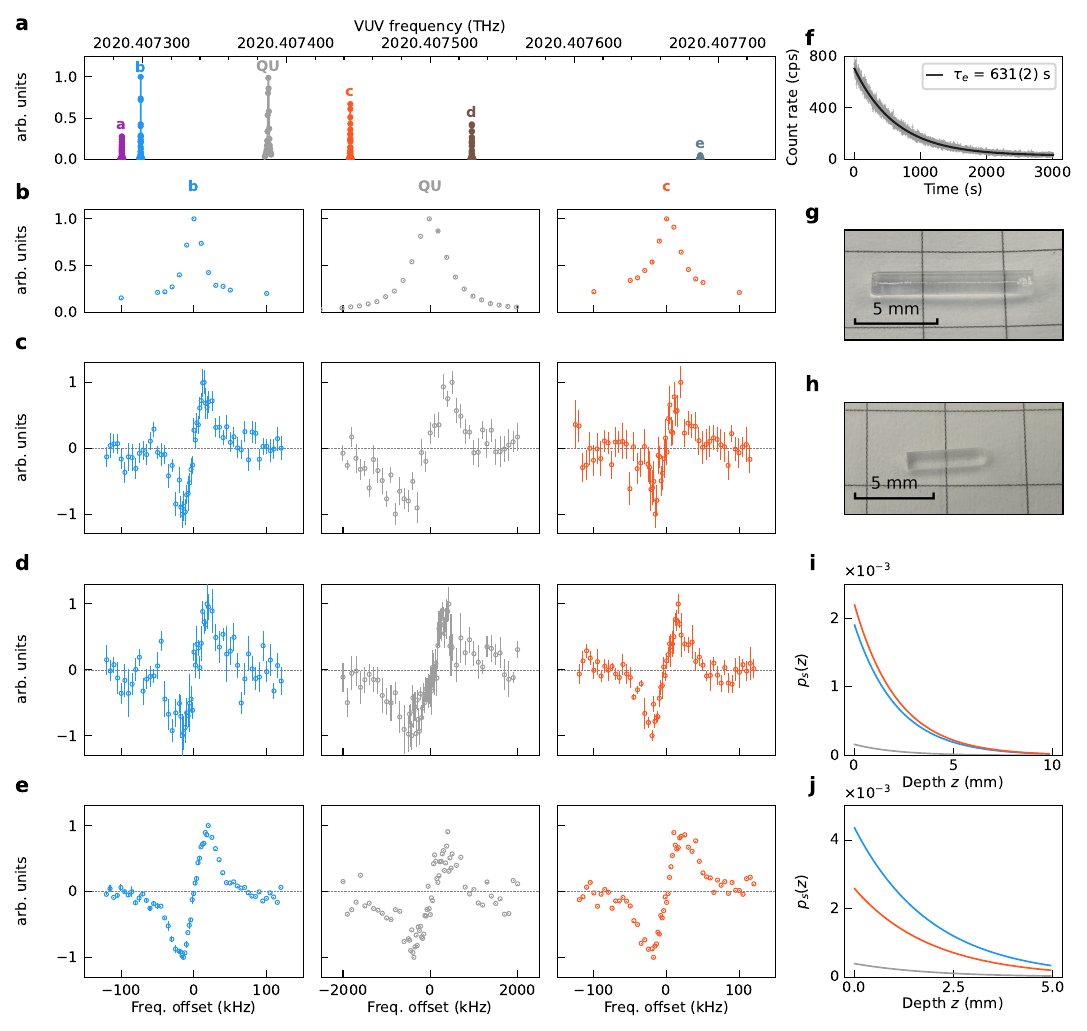}
    \caption{
    \textbf{Fluorescence and absorption spectroscopy of $^{229}$Th:CaF$_2$ crystals.}
    \textbf{a,} Spectral overview of the nuclear transitions measured in TS1 on an absolute VUV-frequency axis. 
    {The fitted QR and QU profiles are shown,
    with all data points normalized to the maximum measured signal in the spectrum.
    }
    \textbf{b,} {Expanded, normalized fluorescence excitation spectra of line b (left, blue), QU (centre, grey) and line c (right, orange) from \textbf{a}.} 
    \textbf{c,} Frequency-modulation absorption signals of the same three spectral components measured with an analogue-current PMT readout in S1. Each data point is acquired for 80~s.  
    \textbf{d,} Corresponding absorption signals measured with the phototube in S1. Each data point is acquired for 40~s. 
    \textbf{e,} Absorption signals measured with the phototube in TS1. Each data point is acquired for 10~s.  
    In \textbf{b--e}, the frequency offset is referenced to the fitted centre of each spectral component, and the signals are normalized. 
    Separate fine absorption scans across QU, performed with reduced modulation amplitudes, reveal no reproducible narrow features on frequency scales of approximately \(500~\mathrm{Hz}\) or larger.
    \textbf{f,} Time-resolved nuclear fluorescence decay measured in TS1 following VUV excitation on QU, fitted with a single-exponential decay. 
    \textbf{g,} S1 crystal.
    \textbf{h,} TS1 crystal.
    {\textbf{i,j}, Calculated depth-dependent steady-state excitation fraction, $p_s(z)$, for resonant excitation through line b (blue), QU (grey), and line c (orange) in S1 (\textbf{i}) and TS1 (\textbf{j}). Here, $z$ is measured from the VUV entrance surface. The calculation uses the transition line strengths, measured ensemble linewidths and nuclear lifetime, together with the local VUV photon flux after accounting for surface reflection and off-resonant propagation loss through the crystal (Methods). a.u., arbitrary units.
    }}
    \label{fig:2}
\end{figure*}

We generate continuous-wave 148.4-nm radiation by resonance-enhanced four-wave mixing in cadmium vapour~\cite{DingVUV2025,DingVUVproposal2024,Mat_elem}. 
{The 375-nm field is tuned to the cadmium $5^1S_0 \rightarrow 6^1S_0$ two-photon resonance, which enhances the third-order non-linear response, and a 710-nm field participates in the sum-frequency four-wave-mixing process to generate VUV radiation at 148.4~nm.}
The 375-nm light is generated by second-harmonic generation of a fibre-based 750-nm laser in a bow-tie enhancement cavity, delivering 5 W of ultraviolet power. Together with 10 W of 710-nm light from another fibre-based laser system, the two fundamental beams are combined and sent into a cadmium oven operated at 600 $^{\circ} \mathrm{C}$, where 10 $\mu$W of continuous-wave 148.4-nm radiation is generated. 
After the oven, the VUV light is spatially separated from the fundamental beams using a Brewster-angle MgF$_2$ prism~\cite{DingVUV2025} and approximately 5 $\mu$W of VUV power is delivered to the front surface of the $^{229}$Th:CaF$_2$ crystal {with a Gaussian diameter of approximately 1~mm} (Fig.~\ref{fig:1}a,b).

The 750-nm and 710-nm lasers are phase-locked to modes of a self-referenced femtosecond Er-fibre frequency comb, whose carrier-envelope-offset frequency, $f_{\rm CEO}$, is stabilized with an $f$--$2f$ interferometer to a hydrogen maser-referenced direct digital synthesizer~\cite{lin202187sr}. 
The repetition rate, $f_{\rm rep}$, is stabilized by locking the 1,397-nm comb branch to an ultrastable 1,397-nm laser referenced to a cryogenic silicon cavity with $10^{-17}$ level stability~\cite{chen2025laser}. 
The comb therefore transfers the stability of the silicon cavity to the 710-nm and 750-nm lasers, while $f_{\rm rep}$ and $f_{\rm CEO}$ are counted against the hydrogen maser to provide traceability of the VUV frequency. Under the operating conditions used for spectroscopy and clock operation, the VUV linewidth is estimated to be sub-Hertz, well below the observed inhomogeneous linewidths of the nuclear transitions in the crystal (Fig.~\ref{fig:1}c).

\vspace{0.6cm}
\noindent \textbf{Fabrication of $^{229}$Th:CaF$_2$ crystals}

Two $^{229}$Th:CaF$_2$ crystals, S1 (Fig.~\ref{fig:2}g) and TS1 (Fig.~\ref{fig:2}h), are fabricated for continuous-wave laser nuclear spectroscopy and clock operation. Both crystals are prepared as elongated cylinders, providing a long VUV interaction path and hence sufficient \(^{229}\)Th column density for absorption readout.
S1 is grown at the Shanghai Institute of Optics and Fine Mechanics using the Bridgman--Stockbarger method, with medium-frequency induction heating.
S1 has a diameter of 1.44~mm, a length of 9.8~mm and an average $^{229}$Th concentration of $9.7\times10^{16}$~cm$^{-3}$.
{The crystal contains small bubbles along its central axis (Fig.~\ref{fig:2}g) and has a VUV transmission of approximately 1\%.
The limited VUV transmission and $^{229}$Th concentration, together with the small fraction of incorporated nuclei occupying the QR site~\cite{Hiraki2025LaserMossbauer}, reduce the signal-to-noise ratio (SNR) of the absorption discriminator.}
Nevertheless, S1 still supports fluorescence spectroscopy, absorption spectroscopy and initial nuclear clock operation.

To address these limitations and test the reproducibility of the clock-transition frequency across crystals, TS1 is fabricated using the temperature-gradient technique~\cite{SuTGT} through a collaboration between the Shanghai Institute of Ceramics of Chinese Academy of Sciences and Tsinghua University. {This method employs a vertical temperature gradient, as in Bridgman--Stockbarger growth, but keeps the crucible stationary. Crystallization is driven by gradually lowering the temperature of the graphite heater, rather than by translating the crucible, with a cooling rate of \(2~^{\circ}\mathrm{C}\,\mathrm{h}^{-1}\) and a pressure below \(1\times10^{-3}~\mathrm{Pa}\).}

Using this approach, we fabricate TS1 from a growth solution containing only 1.4~$\mu$g (10~kBq) of $^{229}$Th.  {The small isotope amount, dictated by the severe scarcity of $^{229}$Th, leaves essentially no room for iterative growth optimization.}
The resulting bubble-free TS1 crystal has a diameter of 1.09~mm, a length of 4.94~mm, an average $^{229}$Th concentration of $2.2\times10^{17}$~cm$^{-3}$, and a VUV transmission of 7\%.

\vspace{0.6cm}
\noindent \textbf{Continuous-wave fluorescence spectroscopy}

We perform continuous-wave fluorescence spectroscopy {on both S1 and TS1. At each frequency point, the crystal is illuminated with the VUV laser for 400~s, during which the transmitted VUV power is simultaneously monitored by the phototube. The 148-nm nuclear fluorescence is then recorded with the photomultiplier tube (PMT) for 200~s. 
The fluorescence-signal analysis accounts for variations in the excitation power and for residual population from preceding measurements due to the comparable nuclear excited-state lifetime and measurement-cycle duration (Methods).
Both crystals exhibit a broad quadrupole-unresolved (QU) structure, consistent with a small effective electric-field gradient~\cite{Hiraki2025LaserMossbauer}, together with a quadrupole-resolved (QR) manifold in which the nuclear transition is split into five narrow components~\cite{LaserSpecYe,Hiraki2025LaserMossbauer} (Fig.~\ref{fig:1}d).
The spectra shown in Fig.~\ref{fig:2}a,b are recorded in TS1.}

{Within the QR manifold, the $m_{\mathrm{g}}=\pm5/2 \rightarrow m_{\mathrm{is}}=\pm3/2$ and $m_{\mathrm{g}}=\pm1/2 \rightarrow m_{\mathrm{is}}=\pm1/2$ transitions are labelled lines b and c, respectively, following ref.~\cite{LaserSpecYe}. 
In TS1, lines b and c have fitted full-width at half-maximum (FWHM) values of $27(1)~\mathrm{kHz}$ and $39(2)~\mathrm{kHz}$, respectively, consistent with those previously reported for low-$^{229}$Th-concentration crystals~\cite{FrequencyReproducibility2025} (Fig.~\ref{fig:2}b; Methods).
For the QU structure, the fitted linewidth is \(0.898(5)~\mathrm{MHz}\). 
Following resonant excitation of QU, we record the subsequent fluorescence decay and obtain an excited-state lifetime of $631(2)~\mathrm{s}$ from a single-exponential fit (Fig.~\ref{fig:2}f).}

\vspace{0.6cm}
\noindent \textbf{Absorption spectroscopy}

To obtain a rapid spectroscopic readout of the long-lived nuclear excitation, we monitor the VUV power transmitted through the crystal and measure the nuclear response in absorption. Because the on-resonance fractional absorption of the VUV light is small (approximately $10^{-4}$),  we modulate the VUV frequency sinusoidally and demodulate the transmitted VUV signal to extract the absorption-induced response. Normalizing the demodulated signal to the simultaneously recorded direct current transmission produces a dispersive-like signal (Fig.~\ref{fig:1}e), which is used as the frequency discriminator for clock operation. Fitting the full frequency-modulated line shape yields the linewidth of the underlying nuclear transition and its on-resonance fractional absorption of the VUV light (Methods).

{
We characterize the absorption readout performance by the 1-s absorption SNR, defined as the peak of the normalized demodulated absorption signal relative to the baseline noise projected to a 1-s measurement time (Methods).
In the photon-shot-noise limit, the 1-s absorption SNR scales as \(\mathrm{SNR}_{1\mathrm{s}}
\propto A[\dot{N}_e\cdot(1~\mathrm{s})]^{1/2}\),
where $A$ is the on-resonance fractional VUV absorption and
$\dot{N}_e$ is the detected photoelectron rate (Methods).
This scaling motivates the successive optimization of the photodetection scheme and crystal properties described below.
}

We first perform absorption spectroscopy on S1 using a PMT operated in analogue-current mode (Fig.~\ref{fig:2}c). {Lines b and c, together with QU, are resolved after sufficiently long measurement times. Although dynode multiplication provides sufficient gain for detecting the weak absorption signal, PMT saturation limits the VUV power reaching the detector to approximately $3~\mathrm{nW}$. The resulting low detected photoelectron rate leads to a large relative photon shot noise, limiting the 1-s absorption SNR.}

We next replace the PMT with a phototube operated in direct-photocurrent mode, which {accommodates substantially greater VUV power without saturation. The photocurrent is amplified with a low-noise transimpedance amplifier, enabling a higher detected photoelectron rate without a substantial electronic-noise penalty.
With S1 unchanged, this detector upgrade improves the 1-s absorption SNR
(Fig.~\ref{fig:2}d). }

{
In the third stage, we retain the phototube readout and replace S1 with TS1.
The higher VUV transmission of TS1 increases the transmitted VUV power to approximately \(200~\mathrm{nW}\) and the detected photoelectron rate by approximately 30-fold relative to the S1 + PMT configuration, after correcting for detector quantum efficiencies.
Together with the approximately threefold larger measured b-line fractional absorption in TS1, this raises the b-line 1-s absorption SNR to approximately 20 times that obtained with the S1 + PMT configuration (Fig.~\ref{fig:2}e).}

{
The increase in 1-s absorption SNR across the three configurations allows the laser detuning from the nuclear resonance to be determined more precisely. 
We next use the normalized demodulated absorption signal in each configuration for nuclear clock operation and observe a corresponding improvement in the fractional frequency instability.
}

\vspace{0.6cm}
\noindent \textbf{Nuclear clock operation and instability}

\begin{figure}[htbp]
    \centering
    \includegraphics[width=\linewidth]{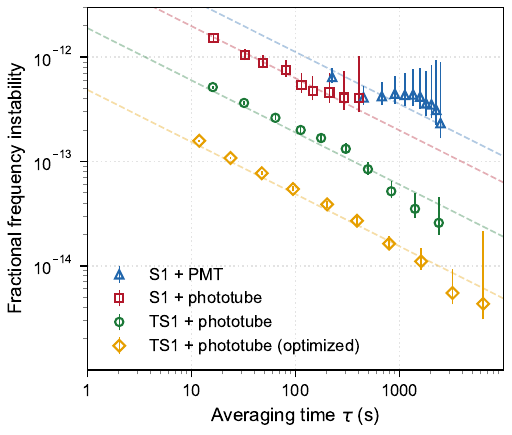}
    \caption{{\textbf{Fractional frequency instability of the nuclear clock.}
    Overlapping Allan deviations measured for four configurations:
    S1 + PMT, S1 + phototube, 
    TS1 + phototube and optimized TS1 + phototube.
    Dashed lines, weighted $\tau^{-1/2}$ fits.
    Error bars, 68.3\% confidence intervals calculated from the
    equivalent degrees of freedom of the overlapping Allan-deviation estimator under the assumed white-frequency-noise model.
    The modulation frequencies, modulation amplitudes, lock-in time
    constants, measurement times, and feedback-cycle durations used for
    the four instability measurements are summarized in Extended Data
    Table~\ref{tab:measurement_parameters}.
    }}
    \label{fig:3}
\end{figure}

We operate the nuclear clock by locking the comb-referenced 148.4-nm VUV laser to line b, which combines a narrow linewidth with weak temperature sensitivity~\cite{Temperature,FrequencyReproducibility2025}. In each interrogation cycle, the normalized demodulated absorption signal is converted to a residual detuning using the calibrated discriminator slope.  This detuning is fed back through an acousto-optic modulator (AOM) in the comb-referenced optical chain (Fig.~\ref{fig:1}c) to update the VUV interrogation frequency for the next cycle. The resulting cycle-by-cycle VUV frequency record is used to calculate the in-loop fractional frequency instability shown in Fig.~\ref{fig:3}.
Between nuclear interrogations, the VUV laser inherits the stability of the cryogenic-silicon-cavity-stabilized optical reference through the comb-referenced optical chain. The independently characterized fractional instability of this reference remains below \(10^{-15}\) at \(1,000~\mathrm{s}\), well below the measured nuclear clock instability at comparable averaging times~\cite{chen2025laser}. This supports using the in-loop record as a faithful estimate of the instability of the operating nuclear clock.

{{
Using this interrogation-and-feedback scheme, the three absorption-readout configurations described above yield progressively improved clock instability (Fig.~\ref{fig:3}). The fitted fractional frequency instabilities decrease from
$1\times10^{-11}/\sqrt{\tau/\mathrm{s}}$ for S1 + PMT to
$6\times10^{-12}/\sqrt{\tau/\mathrm{s}}$ for S1 + phototube and
$2\times10^{-12}/\sqrt{\tau/\mathrm{s}}$ for TS1 + phototube before further optimization.
}}

We then optimize the TS1 + phototube configuration by increasing the average transmitted VUV power to 350 nW, increasing the measurement duty cycle to approximately 85\%, reducing residual amplitude modulation, and using a near-optimal modulation amplitude of 12~kHz (Fig.~\ref{fig:1}e and Extended Data Table~\ref{tab:measurement_parameters}; Methods). 
The remaining \(15\%\) dead time arises from technical overhead in the present digital-feedback implementation and could be substantially reduced through improved feedback control.
The overlapping Allan deviation is statistically consistent with
$\tau^{-1/2}$ scaling over the measured interval, with a fitted
instability of
\begin{equation}
\sigma_y(\tau)
=
5\times10^{-13}
/\sqrt{\tau/\mathrm{s}} .
\end{equation}
This instability is within a factor of two of the photon-shot-noise limit calculated from the measured absorption-discriminator parameters (Methods).

{In all four configurations, the crystal temperature and magnetic field are controlled to limit the systematic shifts and resulting frequency fluctuations, which affect the inter-crystal frequency comparisons and clock instability, respectively.
The measured crystal temperature is $301.4(1)~\mathrm{K}$,  with a systematic standard uncertainty of $0.1~\mathrm{K}$ dominated by the temperature-sensor calibration.
For line b, the root-mean-square (RMS) temporal temperature fluctuation of $3~\mathrm{mK}$ corresponds to a frequency fluctuation of approximately $6~\mathrm{Hz}$, or a fractional frequency fluctuation of approximately $3\times10^{-15}$~(refs. ~\cite{Temperature,FrequencyReproducibility2025}). 
The ambient magnetic field is suppressed by multilayer $\mu$-metal shielding, while a bifilar heater winding and twisted current leads minimize additional current-induced fields. The residual magnetic field is estimated by simulation to be approximately $1~\mathrm{mG}$, corresponding to an estimated Zeeman contribution below $1~\mathrm{Hz}$~\cite{CrystalSchumm,FrequencyReproducibility2025}.}

To test the inter-crystal reproducibility of the clock transition, we compare the centre frequencies of line b in S1 and TS1, obtaining $\nu_{b,\mathrm{S1}}=2{,}020{,}407{,}298{,}798.077(63)_{\mathrm{stat}}(185)_{\mathrm{sys}}~\mathrm{kHz}$ and $\nu_{b,\mathrm{TS1}}=2{,}020{,}407{,}298{,}798.635(20)_{\mathrm{stat}}(185)_{\mathrm{sys}}~\mathrm{kHz}$, respectively (Methods). For this inter-crystal comparison, the relevant temperature contribution is set by the residual relative-temperature uncertainty. Combining this contribution with the statistical line-centre uncertainties, we obtain an inter-crystal frequency difference of \textcolor{black}{$5.6(13)\times10^2$}~Hz,
corresponding to a fractional frequency difference of $2.8(6)\times10^{-13}$.

{
For an inter-laboratory comparison, we evaluate the quadratic fit to the published JILA C10 and C13 b-line data~\cite{FrequencyReproducibility2025} at \(301.4(1)~\mathrm{K}\), obtaining an inferred frequency of
\textcolor{black}{\(\nu_{b,\mathrm{JILA}}=2{,}020{,}407{,}298{,}798.58(122)~\mathrm{kHz}\)},   
where the uncertainty is a combined standard uncertainty dominated by the JILA temperature-dependent fit (Methods). 
The frequency offsets of S1 and TS1 relative to this inferred JILA value are \textcolor{black}{\(-0.50(122)~\mathrm{kHz}\) and \(+0.06(122)~\mathrm{kHz}\)}, respectively, both consistent with zero within the combined standard uncertainties.
These comparisons therefore provide a quantitative assessment of the b-line frequency reproducibility across independently fabricated crystals and independent experimental platforms. }

\vspace{0.6cm}
\noindent \textbf{Conclusion}

We perform continuous-wave laser spectroscopy of the $^{229}$Th isomeric transition in home-grown $^{229}$Th:CaF$_2$ crystals and resolve {both QU and QR in fluorescence and absorption spectroscopy.}
We demonstrate an operating nuclear clock based on high-bandwidth, high-SNR phototube absorption readout. The phototube detection is compatible with substantially higher transmitted VUV power than PMT readout, thereby overcoming the photon flux bottleneck and enabling robust absorption locking.
The clock reaches a fractional frequency instability of {$5\times10^{-13}/\sqrt{\tau/s}$, averaging down to the $10^{-15}$ level over the measured interval.}
Temperature fluctuations and the residual Zeeman contribution are below the present instability limit.
Together with the crystal-to-crystal agreement of the clock-transition frequency, these results establish laser-addressed nuclei as operational clock references and \(^{229}\)Th:CaF\(_2\) as a stable, reproducible solid-state platform for compact nuclear clocks.

\vspace{0.6cm}
\noindent \textbf{Acknowledgements.}
We thank J. Ye and C. Zhang for discussions, R. Yu, X. Feng, J. Yuan, F. Tang and R.-Q. Lei for assistance. S.D. thanks Q. Xue, W. Duan and H. Zhai for their support and co-ordination with crystal fabrication. 

\vspace{0.6cm}
\noindent \textbf{Funding.}
This work is supported by Tsinghua University Dushi Program, Beijing Academy of Quantum Information Sciences (grant no. BAQISY25NG01), Beijing Science and Technology Planning Project (grant no. Z25110100040000), and the National Natural Science Foundation of China (grant nos. 12341401 and 12274253); SIOM group is supported by Strategic Priority Research Program of the Chinese Academy of Sciences (grant no. XDB0920000), National Natural Science Foundation of China (grant nos. 12341404 and 12341402), and Zhangjiang Laboratory (grant no. ZJSP21A001D); SICCAS group is supported by CAS Project for Young Scientists in Basic Research (grant no. YSBR-024); X.Z. acknowledges support from the Strategic Priority Research Program of the Chinese Academy of Sciences (grant no. XDB35020100) and the Quantum Science and Technology-National Science and Technology Major Project (grant no. 2021ZD0301903).

\vspace{0.6cm}
\noindent \textbf{Author contributions.}
B.H., G.Y., Q.X., W.B., G.P., Z.~Zhan, L.~Yan, Y.~Wang, N.M., J.~Li, Y.~Wu and S.D. constructed the experimental setup and carried out the measurements;
B.H., Q.X., Z.~Zhang, C.Y., Z.~Zhan, D.J., X.Q., X.~Liu, Q.H., T.S., J.~Lin, L.S. and S.D. fabricated and characterized the TS1 crystal; C.Z., P.Z., L.L., S.L., Q.G., Y.~Li, L.M., S.Z. and Y.H. fabricated and characterized the S1 crystal;
Z.-A.C. and X.Z. provided the ultrastable 1,397-nm laser referenced to the cryogenic silicon cavity;
H.T., B.L. and Y.~Lin provided the hydrogen maser;
H.S., X.~Li, Y.M. and L.~You provided resources and contributed to discussions;
S.D. conceived, initiated and supervised the project; B.H., Q.X. and S.D. wrote the manuscript with contributions from L.~Yan and input from all authors.

\vspace{0.6cm}
\noindent \textbf{Competing interests.}
The authors declare no competing interests.

\vspace{0.6cm}
\noindent \textbf{Correspondence and requests for materials} should be addressed to Wenhao Bu, Yin Hang, Liangbi Su and Shiqian Ding.

\setcounter{firstbib}{\value{NAT@ctr}}

\makeatletter
\def\endthebibliography{%
    \def\@noitemerr{\@latex@warning{Empty `thebibliography' environment}}%
    \endlist
}
\makeatother
\section{Methods}

\vspace{0.6cm}
\noindent\textbf{Fluorescence-signal analysis}

The fluorescence spectra of S1 and TS1 were obtained by correcting for two effects in the measured fluorescence signals. Residual excited-state population from preceding measurements was removed using a procedure adapted from
refs.~\cite{LaserSpecSchumm,FrequencyReproducibility2025}, whereas variations in the excitation were corrected using the monitored transmitted VUV power during each excitation interval.

For each spectrum, the fluorescence records from all frequency points $i$ are fitted jointly to
\begin{equation}
C_i(t)
=
B+A_i\exp(-t/\tau_e),
\end{equation}
where \(t\) is measured from the beginning of the acquisition window,
\(A_i\) is the fluorescence amplitude at the beginning of that window,
and \(B\) is a common constant background count rate for the
corresponding spectrum. The lifetime is fixed to the independently
measured value \(\tau_e=631~\mathrm{s}\).

Because fluorescence acquisition of each frequency point begins before the residual excited-state population from previous acquisitions has decayed fully, the measured fluorescence records contain contributions from earlier excitations. The net fluorescence amplitude at each frequency point is therefore obtained by subtracting the preceding-point amplitude extrapolated to the start of the current fluorescence-acquisition window:

\begin{equation}
A_{i,\mathrm{net}}
=
A_i
-
A_{i-1}
\exp\left[
-\frac{
t_{i,\mathrm{det}}-t_{i-1,\mathrm{det}}
}{
\tau_e
}
\right],
\end{equation}
where \(t_{i,\mathrm{det}}\) and \(t_{i-1,\mathrm{det}}\) are the
start times of the fluorescence-acquisition
windows of the current and previous frequency points, respectively. 

To correct for variations in the excitation arising from VUV power fluctuations, the transmitted VUV power is monitored during each excitation interval.
In the weak-excitation regime, the excited-state population generated
during an infinitesimal excitation interval of $dt$ is proportional to the instantaneous transmitted VUV
power $P(t)$, whereas population excited at an earlier time is reduced by
spontaneous decay before the fluorescence-acquisition window
~\cite{CrystalSchumm,LarsCalculation}. The effective excitation dose for
the frequency point $i$ is therefore
\begin{equation}
D_i
=
\int_{t_{i,\mathrm{on}}}^{t_{i,\mathrm{off}}}
P(t)
\exp\left[
-\frac{t_{i,\mathrm{det}}-t}{\tau_e}
\right]dt ,
\end{equation}
where $t_{i,\mathrm{on}}$ and $t_{i,\mathrm{off}}$ denote the start and end of the excitation interval of frequency point $i$.
 
The net fluorescence amplitude is then normalized to the effective excitation dose to obtain a corrected fluorescence response:
\begin{equation}
F_i
=
\frac{
A_{i,\mathrm{net}}
}{
D_i
}.
\end{equation}

This normalization makes the corrected fluorescence response insensitive to the frequency point acquisition order and variations in the VUV excitation.

The corrected fluorescence responses $F_i$ constitute the fluorescence spectrum for each spectral component $s$. The spectrum is fitted to a single-Lorentzian profile
\begin{equation}
F_s(\nu)
=
F_{0,s}
+
\frac{H_s}
{1+4[(\nu-\nu_{0,s})/\Delta\nu_s]^2},
\end{equation}
where \(F_{0,s}\) is the constant offset, \(H_s\) is the Lorentzian peak height, \(\nu_{0,s}\) is the line-centre
frequency, and \(\Delta\nu_s\) is the FWHM. 
Applying this Lorentzian fit to TS1 yields linewidths of
$\Delta\nu_b=27(1)~\mathrm{kHz}$,
$\Delta\nu_c=39(2)~\mathrm{kHz}$, and
$\Delta\nu_{\mathrm{QU}}=0.898(5)~\mathrm{MHz}$.

For the narrow QR transitions, the spectra of lines a--d in both S1 and TS1 show evidence of a broad pedestal in addition to the narrow central component.
A representative example, line b in TS1, is shown in Extended Data
Fig.~\ref{fig:ED3_pedestal}. 
The pedestal could reflect an inhomogeneously broadened subensemble of QR sites, arising from variations in the local fluoride environment, lattice strain or nearby defects, although its microscopic origin remains to be established.

\vspace{0.6cm}
{
\noindent\textbf{Absorption signal analysis}

To extract transition linewidths and on-resonance fractional absorptions, we fit the demodulated absorption signals to the first-harmonic response of a sinusoidally frequency-modulated Lorentzian line shape.

For each spectral component \(s\in\{b,c,\mathrm{QU}\}\), the measured normalized demodulated absorption signal is
\begin{equation} S_{1f,s}^{\mathrm{RMS}}(\bar{\nu}) \equiv \frac{V_{1f,s}^{\mathrm{RMS}}(\bar{\nu}) }{ V_{\mathrm{DC}}(\bar{\nu}) }, \label{eq:normalized_1f_rms} 
\end{equation}
where $V_{1f,s}^{\mathrm{RMS}}(\bar{\nu})$ is the in-phase first-harmonic voltage recorded by the lock-in amplifier in the root-mean-square units, and  $V_{\mathrm{DC}}(\bar{\nu})$ is the direct current signal of the simultaneously monitored transmitted VUV light.

The normalized demodulated absorption signal is modelled as
\begin{equation}
S^{\mathrm{RMS}}_{1f,s}(\bar{\nu})
=
\frac{1}{\sqrt{2}}A_s\mathcal{D}_s(\bar{\nu})+E_s,
\label{Eq::absorption_fit}
\end{equation}
where
$A_s
\equiv
(P_{\mathrm{off}}-P_{\mathrm{on},s})/P_{\mathrm{off}}
$
is the on-resonance fractional absorption of the VUV light, with $P_{\mathrm{off}}$ the off-resonant transmitted power and $P_{\mathrm{on},s}$ the transmitted power
at the centre of spectral component $s$ without frequency modulation. $E_s$ is a fitted constant offset. The function \(\mathcal{D}_s(\bar{\nu})\) is the first-harmonic response function for a unit-height Lorentzian absorption profile $\mathcal{L}_s(\nu)$, 
\begin{equation}
\mathcal{D}_s(\bar{\nu})
=
\frac{1}{\pi}
\int_0^{2\pi}
\mathcal{L}_s
\bigl(
\bar{\nu}+\delta_s\cos\theta
\bigr)
\cos\theta\,d\theta,
\end{equation}
where
\begin{equation}
\mathcal{L}_s(\nu)
=
\left[
1+
4\left(
\frac{\nu-\nu_{0,s}}
{\Delta\nu_s}
\right)^2
\right]^{-1}\label{eq:lorentzian_profile}.
\end{equation}

Here, \(\bar{\nu}\) is the mean laser frequency, \(\delta_s\) is the modulation amplitude, \(\theta\) is the modulation phase, \(\nu_{0,s}\) is the line-centre frequency, and \(\Delta\nu_s\) is the Lorentzian FWHM. Fits of \eqref{Eq::absorption_fit} to the measured normalized demodulated signals yield the on-resonance fractional absorptions $A_s$ and transition linewidths $\Delta\nu_s$.

The 1-s absorption SNR in the main text is defined as the peak amplitude of the normalized demodulated absorption signal divided by the standard deviation of the signal baseline projected to a 1-second measurement time: 
\begin{equation}
\begin{aligned}
\mathrm{SNR}_{1\mathrm{s}}
&\equiv
\frac{
\displaystyle
\max_{\bar{\nu}}
\left|S^{\mathrm{RMS}}_{1f,s}
\right|
}{
\displaystyle
\sigma_{1\mathrm{s}}
\left[
S^{\mathrm{RMS}}_{1f,s}
\right]
}.
\\
\end{aligned}
\end{equation}

In the photon-shot-noise limit, it reduces to
\begin{equation}
\begin{aligned}
\mathrm{SNR}_{1\mathrm{s,SN}}
&\simeq 
\frac{
\displaystyle
A_s\max_{\bar{\nu}}
\left|
\mathcal{D}_{s}(\bar{\nu})
\right|/\sqrt{2}
}{
\displaystyle
\sqrt{1/[\dot{N_e}\cdot(1~\mathrm{s})]}
}
\\
&=
k_s A_s\sqrt{\dot{N}_e\cdot(1~\mathrm{s})},
\end{aligned}
\label{eq:absorption_snr_derivation}
\end{equation}
where $\dot{N_e}$ is the detected photoelectron rate and $k_s=
\frac{1}{\sqrt{2}}
\max_{\bar{\nu}}
\left|
\mathcal{D}_s(\bar{\nu})
\right|$ is the proportionality factor for the spectral component $s$. The dependence of the $\mathrm{SNR}_{1\mathrm{s,SN}}$ on both $A_s$ and $\dot{N}_e$ motivates the successive improvements to the detector and crystal used in absorption spectroscopy and clock operation.
}

\vspace{0.6cm}

\noindent\textbf{Resonant steady-state excitation fraction}

To quantify the resonant nuclear excitation throughout the crystal, we calculate the depth-dependent steady-state excitation fraction $p_s(z)$ for the laser tuned to the centre of each spectral component, $s\in\{b,c,\mathrm{QU}\}$, following ref.~\cite{LarsCalculation}. In the weak-excitation regime,
\begin{equation}
p_s(z)
=
\Gamma_{\mathrm{exc}}^{(s)}(z)\tau_e,
\label{eq:population_fraction}
\end{equation}
where \(\Gamma_{\mathrm{exc}}^{(s)}(z)\) is the excitation rate and
\(\tau_e\) is the measured nuclear lifetime.

The Rabi frequency for component \(s\) is
\begin{equation}
\Omega_s(z)
=
\left[
\frac{2\pi c^2 I(z)\chi_s\Gamma_\gamma}
{\hbar\omega_0^3}
\right]^{1/2},
\end{equation}
where \(I(z)\) is the local VUV intensity, \(\omega_0\) is the angular
nuclear-transition frequency,
\(\Gamma_\gamma\simeq2\pi\times63~\mu\mathrm{Hz}\) is the radiative decay rate calculated from the measured excited-state lifetime, and \(\chi_s\) is the effective line-strength factor. 
For $s=b,c$ and $\mathrm{QU}$, the effective line-strength factors are, respectively,
\begin{equation}
\left(
\chi_b,\chi_c,\chi_{\mathrm{QU}}
\right)
=
\left(
\frac{2}{9},
\frac{1}{5},
\frac{2}{3}
\right).
\end{equation}
These values are obtained by summing the squared Clebsch--Gordan coefficients over the unresolved magnetic-sublevel transitions and averaging the resulting coupling strengths over the orientations of the effective quantization axes relative to the laser polarization~\cite{CrystalSchumm}.

Because resonant nuclear absorption is negligible compared with the
off-resonant propagation loss, the local VUV intensity is approximated as
\begin{equation}
I(z)
\simeq
\frac{P_{\mathrm{in}}(1-R)}
{S_{\mathrm{beam}}}
\exp(-\alpha_{\mathrm{nr}}z),
\end{equation}
where \(P_{\mathrm{in}}\) is the incident VUV power and
\(S_{\mathrm{beam}}\) is the effective illuminated area. The Fresnel
reflectance and effective off-resonant attenuation coefficient are
\begin{equation}
\begin{aligned}
R
&=
\left(
\frac{n-1}{n+1}
\right)^2,\\
\alpha_{\mathrm{nr}}
&=
-\frac{1}{L}
\ln\left[
\frac{T_{\mathrm{xtal}}}{(1-R)^2}
\right],
\end{aligned}
\end{equation}
where \(n\) is the refractive index of CaF$_2$, \(L\) is the crystal
length, and \(T_{\mathrm{xtal}}\) is the measured off-resonant transmission
through the crystal. For S1 and TS1, \(T_{\mathrm{xtal}}\) is approximately
\(1\%\) over \(9.8~\mathrm{mm}\) and \(7\%\) over \(4.94~\mathrm{mm}\),
corresponding to equivalent 1-mm internal transmittances of
\(63\%\) and \(60\%\), respectively.

For a spectral component with measured FWHM
\(\Delta\nu_s\), the resonant excitation rate in the weak-excitation
regime is~\cite{LarsCalculation}
\begin{equation}
\Gamma_{\mathrm{exc}}^{(s)}(z)
=
\frac{\Omega_s^2(z)}
{2\pi\Delta\nu_s},
\end{equation}
giving
\begin{equation}
p_s(z)
=
\frac{\Omega_s^2(z)\tau_e}
{2\pi\Delta\nu_s}.
\end{equation}

Using an incident VUV power of approximately $5~\mu\mathrm{W}$ with a Gaussian-beam $1/e^2$ intensity diameter of approximately $1~\mathrm{mm}$, together with the measured linewidths and off-resonant transmissions of S1 and TS1,
we evaluate $p_s(z)$ for the b, c and QU components throughout each crystal.
The resulting depth-dependent excitation fractions are shown in
Fig.~\ref{fig:2}i,j.

For the present experimental parameters, the entrance surface Rabi frequencies are $\Omega_s<2\pi\times1~\mathrm{Hz}$, more than two orders of magnitude smaller than both the crystal-induced optical-decoherence rates and the magnetic-sublevel-mixing rates, which are of order $2\pi\times10^2~\mathrm{Hz}$~\cite{CrystalHudson,CrystalSchumm}. 
This hierarchy justifies the excitation-rate treatment used above, with the measured ensemble linewidth $\Delta\nu_s$ providing the effective linewidth of each spectral component.

\vspace{0.6cm}
{{
\noindent\textbf{Frequency discriminator optimization}

To minimize the measurement-noise contribution to the fractional frequency instability of the nuclear clock, we optimize the modulation amplitude to maximize the slope of the frequency discriminator at the centre of line b. Following the standard stability analysis for atomic clocks~\cite{2015RMP_Optical_atomic_clocks}, the fractional frequency instability arising from statistically independent cycle-to-cycle measurement noise is
\begin{equation}
\sigma_y(\tau)
=
\frac{\sigma_S}
{\nu_0 g_b}
\left(
\frac{T_{\mathrm{c}}}{\tau}
\right)^{1/2},
\label{eq:instability_slope}
\end{equation}
where \(\sigma_S\) is the fluctuation of the normalized demodulated absorption signal at the line centre, \(\nu_0\) is the nuclear-transition frequency, \(T_{\mathrm{c}}\) is the clock-cycle duration, \(\tau\) is the averaging time, and
\begin{equation}
g_b
\equiv
\left|
\frac{\partial S^{\mathrm{RMS}}_{1f,b}(\bar{\nu})}
{\partial\bar{\nu}}
\right|_{\bar{\nu}=\nu_{0,b}}
=
\frac{A_b}{\sqrt{2}}
\left|
\frac{\partial\mathcal{D}_b(\bar{\nu})}
{\partial\bar{\nu}}
\right|_{\bar{\nu}=\nu_{0,b}}
\end{equation}
is the magnitude of the discriminator slope at the line centre. Provided that \(\sigma_S\) is approximately independent of the modulation amplitude over the relevant range, maximizing \(g_b\) minimizes the corresponding contribution to the clock instability.

For a Lorentzian line with FWHM
\(\Delta\nu_b\) and peak modulation amplitude \(\delta_b\),
the discriminator slope is
\begin{equation}
g_b
=
A_b
\frac{4\sqrt{2}\delta_b}{\Delta\nu_b^2}
\left[
1+
4\left(
\frac{\delta_b}{\Delta\nu_b}
\right)^2
\right]^{-3/2}.
\end{equation}
Maximizing \(g_b\) with respect to \(\delta_b\) gives
\begin{equation}
\delta_{b,\mathrm{opt}}
=
\frac{\Delta\nu_b}{2\sqrt{2}}.
\end{equation}
}

For the measured b-line linewidth
$\Delta\nu_b=27(1)~\mathrm{kHz}$, the modulation amplitude that
maximizes the line-centre discriminator slope is
$\delta_{b,\mathrm{opt}}\simeq9.5~\mathrm{kHz}$.
A slightly larger modulation amplitude provides a wider monotonic frequency-discrimination range and hence greater tolerance to
frequency excursions during feedback operation.
We therefore use a near-optimal $\delta_b=12~\mathrm{kHz}$, which increases the width of the monotonic frequency-discrimination range by approximately 13\% while
retaining 96.4\% of the maximum line-centre slope.
At this modulation amplitude, the discriminator slope is
\begin{equation}
    g_{b}=1.05A_b/\Delta\nu_b.
\end{equation}
}

\vspace{0.6cm}
{{
\noindent\textbf{Clock instability at photon-shot-noise limit}

We estimate the photon-shot-noise-limited fractional frequency
instability of the optimized TS1 + phototube configuration from the
detected VUV power. The detected photoelectron rate is
\(\dot{N}_e=\eta_{\mathrm{det}}P_{\mathrm{det}}/(h\nu_0)\), where
\(P_{\mathrm{det}}\) is the detected VUV power and
\(\eta_{\mathrm{det}}\) is the phototube quantum efficiency.
For a measurement time \(T_{\mathrm{m}}\), the photon-shot-noise-limited
standard deviation of the normalized demodulated absorption signal is~\cite{Bjorklund1983FMspectroscopy}
\begin{equation}
\sigma_{S,\mathrm{SN}}
=
\left(
\frac{1}{\dot{N}_e T_{\mathrm{m}}}
\right)^{1/2}.
\label{eq:first_harmonic_shot_noise}
\end{equation}

Substituting this expression into~\eqref{eq:instability_slope} and
using the near-optimized discriminator slope \(g_{b}=1.05A_b/\Delta\nu_b\) gives
\begin{equation}
\sigma_y(\tau)
=
\frac{\Delta\nu_b}
{1.05\,\nu_0 A_b}
\left(
\frac{h\nu_0}
{\eta_{\mathrm{det}}P_{\mathrm{det}}\tau}
\right)^{1/2}
\left(
\frac{T_{\mathrm{c}}}
{T_{\mathrm{m}}}
\right)^{1/2}.
\label{eq:shot_noise_instability}
\end{equation}

For the optimized TS1 + phototube configuration, we use
\(P_{\mathrm{det}}=350~\mathrm{nW}\),
\(\eta_{\mathrm{det}}\simeq0.10\),
\(A_b=2.9\times10^{-4}\),
\(\Delta\nu_b=27~\mathrm{kHz}\),
\(T_{\mathrm{m}}=10.0~\mathrm{s}\), and
\(T_{\mathrm{c}}=11.8~\mathrm{s}\), yielding
\begin{equation}
\sigma_y(\tau)
\simeq
2.9\times10^{-13}
/\sqrt{{\tau}/\mathrm{s}}.
\end{equation}
The measured instability of $5\times10^{-13}/\sqrt{\tau/\mathrm{s}}$
is within a factor of two of this limit, indicating near-photon-shot-noise-limited operation. The remaining difference can be partly accounted for by the gradual decrease in detected VUV power during the measurement.
}}

\vspace{0.6cm}
{\noindent\textbf{Absolute-frequency determination}

For the absolute-frequency measurements of line b in S1 and TS1  at a crystal temperature of \(301.4(1)~\mathrm{K}\), the VUV laser is stabilized to the centre of the corresponding absorption resonance. The locked VUV frequency is determined through the comb-referenced optical chain used for VUV generation, including the comb mode numbers, optical beat notes, AOM shifts and fixed RF offsets. All RF synthesizers and frequency counters are referenced to a common $10$-$\mathrm{MHz}$ hydrogen maser signal.
The reconstructed frequency is corrected for the fractional frequency offset of the hydrogen maser relative to International Atomic Time, using the monthly International Atomic Time clock-rate data for the corresponding measurement periods.

The resulting b-line centre frequencies are
\begin{equation}
\begin{aligned}
\nu_{b,\mathrm{S1}}
&=2{,}020{,}407{,}298{,}798.077(63)_{\mathrm{stat}}(185)_{\mathrm{sys}}~\mathrm{kHz},\\
\nu_{b,\mathrm{TS1}}
&=2{,}020{,}407{,}298{,}798.635(20)_{\mathrm{stat}}(185)_{\mathrm{sys}}~\mathrm{kHz}.
\end{aligned}
\end{equation}
The first set of parentheses gives the $1\sigma$ statistical uncertainty, and the second gives the systematic uncertainty.
These systematic uncertainties arise from the \(0.1~\mathrm{K}\) absolute crystal temperature uncertainty, dominated by the sensor calibration, propagated through the local temperature coefficient of line b reported in ref.~\cite{FrequencyReproducibility2025}.

}

{
\vspace{0.6cm}
\noindent\textbf{Inter-laboratory frequency comparison}

For comparison with the JILA measurements, we fit the publicly available C10 and C13 source data~\cite{ooi_2025_15751158} using the quadratic temperature-dependence model reported in
ref.~\cite{FrequencyReproducibility2025}.
Evaluating the fit at our crystal temperature of
$301.4(1)~\mathrm{K}$ yields
\textcolor{black}{\begin{equation}
\nu_{b,\mathrm{JILA}}
=
2{,}020{,}407{,}298{,}798.58(122)~\mathrm{kHz},
\end{equation}}
where the $1\sigma$ uncertainty combines in quadrature the $1.21~\mathrm{kHz}$ fit uncertainty and the
$0.185~\mathrm{kHz}$ contribution from the uncertainty of our
crystal temperature.

The offsets of the S1 and TS1 measurements from the JILA value are
\textcolor{black}{\begin{equation}
\begin{aligned}
\nu_{b,\mathrm{S1}}-\nu_{b,\mathrm{JILA}}
&=
-0.50(122)~\mathrm{kHz},\\
\nu_{b,\mathrm{TS1}}-\nu_{b,\mathrm{JILA}}
&=
+0.06(122)~\mathrm{kHz}.
\end{aligned}
\end{equation}}
The agreement within the combined standard uncertainties verifies the reproducibility of the b-line frequency across independently fabricated crystals and independent experimental platforms.}

\vspace{0.6cm}
\noindent \textbf{Data Availability}
The datasets generated and analyzed during this study are available in the Figshare repository at https://doi.org/10.6084/m9.figshare.33386191[\cite{Huang2026NuclearClockData}].

\vspace{0.6cm}
\noindent \textbf{Code Availability}
The plotting scripts used to generate the figures from the datasets are available in the Figshare repository at https://doi.org/10.6084/m9.figshare.33386191[\cite{Huang2026NuclearClockData}].


\clearpage

\section*{Extended Data}

\setcounter{figure}{0}
\setcounter{table}{0}

\renewcommand{\thefigure}{\arabic{figure}}
\renewcommand{\thetable}{\arabic{table}}

\makeatletter
\renewcommand{\fnum@figure}{Extended Data Fig.~\thefigure}
\renewcommand{\fnum@table}{Extended Data Table~\thetable}
\makeatother

\begin{figure}[h]
    \centering
    \includegraphics[width=\linewidth]{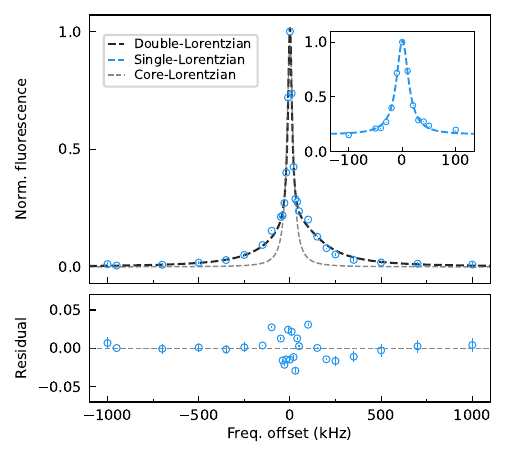}
    \caption{{\textbf{Broad pedestal accompanying line b in TS1.} The main panel shows the full measured spectrum with a phenomenological double-Lorentzian fit over the full range (black dashed line), which captures the narrow component and broad pedestal, and, for comparison, a single-Lorentzian fit to the central data points within $\pm25~\mathrm{kHz}$ of the line centre (grey dashed line).
    Applying the same full-range double-Lorentzian analysis to lines a--d gives narrow-component FWHMs of $75(9)$, $26(1)$, $34(1)$ and $43(3)$~kHz, respectively, and corresponding pedestal FWHMs of $237(31)$, $334(44)$, $292(16)$ and $272(28)$~kHz.
    The inset reproduces the central data shown in Fig.~\ref{fig:2}b of the main text, together with a single-Lorentzian fit with a constant offset (blue dashed line). This fit yields a b-line linewidth of \(27(1)~\mathrm{kHz}\). The lower panel shows the residuals of the full-range double-Lorentzian fit.}}
    \label{fig:ED3_pedestal}
\end{figure}

\clearpage

\begin{table*}[htbp]

\centering
\setlength{\tabcolsep}{2pt}
\renewcommand{\arraystretch}{1.15}
\caption{{\textbf{Frequency-modulation, lock-in-detection and feedback parameters for absorption spectroscopy and clock operation.} Here, $f_{\mathrm{m}}$ is the modulation frequency; $\delta_b$, $\delta_c$ and $\delta_{\mathrm{QU}}$ are the frequency-modulation amplitudes  for line b, line c and the QU structure, respectively; $\tau_{\mathrm{LI}}$ is the lock-in time constant; $T_{\mathrm{m}}$ is the measurement time; $T_{\mathrm{c}}$ is the feedback-cycle duration; and $T_{\mathrm{tot}}$ is the total elapsed duration of the continuous frequency record used for the Allan-deviation analysis. All lock-in measurements use a \(24~\mathrm{dB}\)-per-octave low-pass filter roll-off.}}
\label{tab:measurement_parameters}
\begin{tabular*}{\textwidth}{@{\extracolsep{\fill}}lcccccccc@{}}
\hline
& \multicolumn{4}{c}{Frequency modulation} & \multicolumn{4}{c}{Detection and feedback} \\
\cline{2-5}\cline{6-9}
Configuration & $f_{\mathrm{m}}$ (kHz) & $\delta_b$ (kHz) & $\delta_c$ (kHz) & $\delta_{\mathrm{QU}}$ (kHz) & $\tau_{\mathrm{LI}}$ (s) & $T_{\mathrm{m}}$ (s) & $T_{\mathrm{c}}$ (s) & $T_{\mathrm{tot}}$ (s) \\
\hline
S1 + PMT & 1.49 & 20 & 20 & 250 & 1 & 200 & $\sim225$ & $\sim5{,}400$ \\
S1 + phototube & 2.924 & 20 & 20 & 250 & 0.1 & 10 & $\sim16$ & $\sim1{,}000$ \\
TS1 + phototube & 0.888 & 20 & 20 & 250 & 0.1 & 10 & $\sim16$ & $\sim8{,}800$ \\
\shortstack[l]{TS1 + phototube\\(optimized)} & 0.888 & 12 & \textemdash & \textemdash & 0.1 & 10 & 11.8 & $\sim13{,}600$ \\
\hline
\end{tabular*}
\end{table*}

\end{document}